\documentclass[twocolumn,amssymb,epsfig,aps,lettersize,superscriptaddress,unsortedaddress]{revtex4}
\usepackage{epsfig}
\usepackage{amssymb}
\usepackage{verbatim}

\def\nstruct{$\sim$400}

\usepackage{units}

 \def\kk{{\bf k}}
 
\newcommand{\aflow}{\textsc{Aflow}}
\newcommand{\vasp}{\textsc{vasp}}
\newcommand{\matlab}{\textsc{Matlab}}
\newcommand{\gnuplot}{\textsc{Gnuplot}}
\newcommand{\frozsl}{\textsc{Frozsl}}

\newcommand{\findsym}{\textsc{Findsym}}
\newcommand{\uncle}{\textsc{uncle}}
\newcommand{\apennsy}{\textsc{Apennsy}}
\newcommand{\aconvasp}{\textsc{Aconvasp}}

\begin{document}

\title{AFLOW: an automatic framework for high-throughput materials discovery}
\author{
  Stefano Curtarolo$^{a,b,\star}$, Wahyu Setyawan$^a$, Gus L. W. Hart$^c$, Michal Jahnatek$^a$, Roman  V. Chepulskii$^a$,\\
  Richard H. Taylor$^a$, Shidong Wang$^a$, Junkai Xue$^a$, Kesong Yang$^a$, Ohad Levy$^d$,\\
  Michael J. Mehl$^e$, Harold T. Stokes$^c$, Denis O. Demchenko$^f$, Dane Morgan$^g$}
\address{\small $^a$ Department of Mechanical Engineering and Materials Science, Duke University, Durham, NC 27708}
\address{\small $^b$ Department of Physics, Duke University, Durham, NC 27708}
\address{\small $^c$ Department of Physics and Astronomy, Brigham Young University, Provo, Utah 84602}
\address{\small $^d$ Department of Physics, NRCN, P.O. Box 9001, Beer-Sheva 84190, Israel}
\address{\small $^e$ U.S. Naval Research Laboratory, 4555 Overlook Avenue, SW, Washington DC 20375 }
\address{\small $^f$ Department of Physics, Virginia Commonwealth University, Richmond, VA 23284}
\address{\small $^g$ Department of Materials Science and Engineering, 1509  University Avenue, Madison, WI 53706}
\address{\small $^\star$ Corresponding author email: stefano@duke.edu; tel: +1 919 660 5506; fax: +1 919 6608963}

\begin{abstract}
Recent advances in computational materials science present novel opportunities for structure discovery and optimization, 
including uncovering of unsuspected compounds and metastable structures, electronic structure, surface and nano-particle properties.
The practical realization of these opportunities requires systematic
generation and classification of the relevant computational data by high-throughput methods.
In this paper we present \aflow\  (Automatic Flow), a software framework
for high-throughput calculation of crystal structure properties of
alloys, intermetallics and inorganic compounds. The \aflow\ software
is available for the scientific community on the
website of the materials research consortium, {\sf aflowlib.org}. Its geometric and electronic structure analysis and manipulation tools are
additionally available for online operation at the same website.
The combination of automatic methods and user online interfaces provide a powerful tool for efficient quantum computational materials discovery and characterization. 
\end{abstract}
\maketitle


\section{Introduction}

A new class of software tools to assess material properties has emerged from two parallel theoretical advancements:
quantum-mechanical computations based on density functional theory (DFT), and informatics data mining and evolutionary structure screening strategies.
Joined together, these methods make it possible to efficiently screen large sets of material structures with many
different combinations of elements, compositions and geometrical configurations.
The practical realization of such schemes
necessitates a \emph{high-throughput} (HT) approach, which involves setting up and
performing many {\it ab initio} calculations and then organizing and
analyzing the results with minimal intervention by the user. 
The HT concept has already become an effective and efficient tool for materials 
discovery \cite{Xiang06231995, ceder:nature_1998,Johann02,Stucke03,curtarolo:prl_2003_datamining,morgan:meas_2005_ht,monster,Fischer06} 
and development \cite{koinuma_nmat_review2004,Spivack20035,Boussie2003,Potyrailo2003,Potyrailo2005}.
Examples of computational materials HT applications include
combinatorial discovery of
superconductors \cite{Xiang06231995}, Pareto-optimal search
for alloys and catalysts \cite{johannesson:ref2,johannesson:ref3},
data-mining of quantum calculations applying
principle-component analysis
to uncover new compounds \cite{curtarolo:prl_2003_datamining,morgan:meas_2005_ht,monster,curtarolo:art49,curtarolo:art51,curtarolo:art53,curtarolo:art55,curtarolo:art56,curtarolo:art57,curtarolo:art54,curtarolo:art63,curtarolo:art67,curtarolo:art70},
Kohn-anomalies search in ternary
lithium-borides  \cite{curtarolo:art21,curtarolo:art33,curtarolo:art37},
and multi-optimization techniques used for the study of
high-temperature reactions in multicomponent hydrides
\cite{Wolverton2008,Siegel_PhysRevB.76.134102,Akbarzadeh2007}.

In its practical implementation, the HT approach uses some sort of automatic optimization technique
to screen a library of candidate compounds and to direct further refinements.  The library can be a
set of alloy prototypes \cite{Massalski,monster} or a database of compounds such as the Pauling File
 \cite{Pauling} or the ICSD Database \cite{ICSD,ICSD3}.  Rather than calculating one target physical
quantity over a large number of structures and compositions, a key philosophy of the HT method
is to calculate \emph{a priori} as many different quantities as it is computationally feasible.
Properties and property correlations are then extracted \emph{a posteriori}.

This paper describes the HT framework \aflow\, which we have been
developing over the last decade. The code and manuals describing its
many operation options are freely available for download at {\sf aflowlib.org/aflow.html} \cite{aflowlibPAPER}.
Some of its capabilities may also be operated interactively online at a dedicated web page {\sf aflowlib.org/awrapper.html}.

\section{Software and Structure Database}

The high-throughput framework \aflow, is an assemblage of software tools
comprising over 150,000 lines of C++ code.
It is written for UNIX systems (GNU-Linux, Mac OSX) with the GNU suite of compilers.
Its main features are fully multi-threaded and parallel.
It is designed to run on top of any software for structure energy calculation
and is currently optimized for first-principles
calculations by Vienna Ab Initio Simulation Package ({\small VASP}) \cite{VASP}
(porting to other DFT packages, such as Quantum Espresso
 \cite{quantum_espresso_2009} is underway).
The code offers several options for running the DFT
package, on either a single structure or on sets of structures, by
searching through subfolders containing {\sf aflow.in} input files and running those that have not been
calculated yet.

\aflow\ makes it possible to automatically calculate a suite of physical observables
over a specified class or a large database of structures with little human intervention to set up the input files, run
the calculations, and collate and plot the results. It can also be
used to assist in the setup of standardized calculations of individual
cases. To maximize the usefulness of the large amount of information
produced by the HT approach, the data must be generated
and represented in a consistent and robust manner.
This is especially true when the goal is, as in many materials science applications, to simultaneously
optimize multiple properties. For instance, in catalyst
design \cite{Norskov2009natchem} and superconducting materials
development \cite{curtarolo:art33,curtarolo:art37},
both thermodynamics and electronic structure are essential to the effectiveness of the material.
If started as a common Unix daemon through the queue of a computer cluster, \aflow\
can generate, run, correct and converge many calculations per day,
with minimum human input.

\begin{figure*}
\includegraphics[width=.9\linewidth]{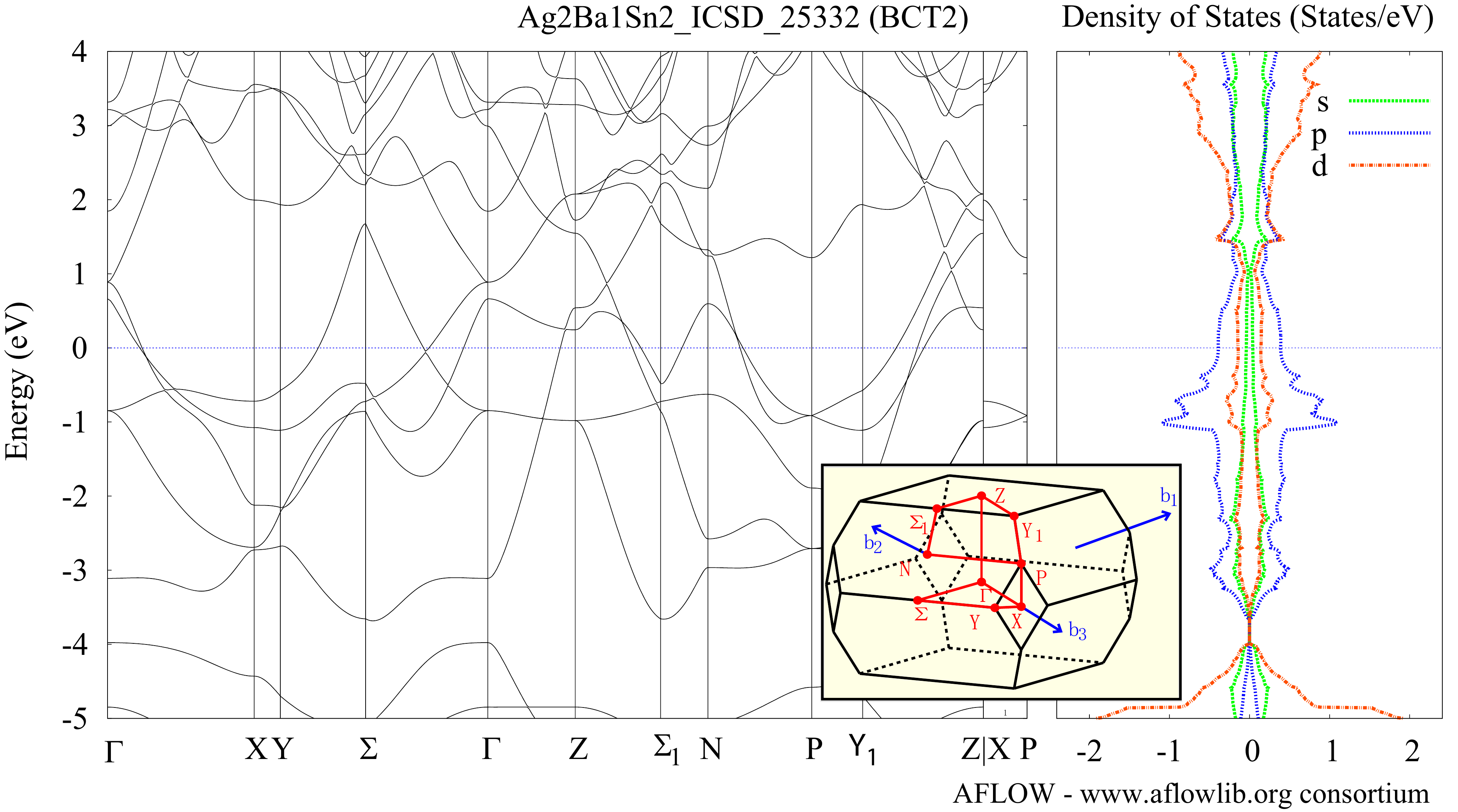}
\caption{ Left: Electronic band structure of
    Ag$_2$BaSn$_2$, ICSD structure \# 25332, calculated along the
    Brillouin Zone path shown in the inset. Right: The electronic
    density of states of the same structure. Both automatically
    generated using the \aflow\ framework. \label{fig1}}
\end{figure*}

The backbone of the software is the implementation of ground-state energy calculations
to identify stable and meta-stable structures.
In addition, it addresses phase diagram construction,
electronic band structure, phonon-spectra and vibrational free-energy, surface stability and adsorption of contaminants on surfaces,
electron-phonon coupling and superconductivity,
local atomic environment expansion \cite{villars:factors}, 
and cluster expansion. The code is being developed continuously with new applications integrated into it periodically.

One of the most difficult challenges in HT database generation is responding
\emph{automatically} to the failure of a calculation. The most common cause is insufficient hardware
resources. Precaution must be taken, for example, to estimate the memory requirement of the tasks
and group jobs so that they do not hamper each other if running on the same
node. The second most common cause of interruption is due to runtime errors of
the {\it ab initio} calculation itself. These errors include inconsistent reciprocal and {\bf k}
lattice meshes, inconsistent atom locations, ill-defined geometries, too few 
 electronic bands to include all the  electrons in the system, internal inconsistencies
arising in the convergence process, and many others.
\aflow\ is capable of detecting most of these problems,
altering the input accordingly and running again. 
This automatic restarting of runs is achieved by diagnosing the error message,
correcting the appropriate parameters, and restarting the calculation.

As of this writing, the \aflow\ database includes \nstruct\ experimental
prototypes taken primarily from the Pauling File \cite{Pauling} with some additions from the
ICSD \cite{ICSD,ICSD3} and the Navy Crystal Lattice database \cite{navy_crystal_prototypes}. The most current list of
these experimental prototypes can be viewed by either using the command-line version of \aflow\ or
the online tool described in Sec.\ \ref{SAT}.
Occasionally, fully relaxed calculations by \aflow\ (cell volume and shape and the basis atom coordinates
inside the cell) turn up as ground state structures that do not yet have known
experimental prototypes \cite{monster,curtarolo:art49,curtarolo:art51,curtarolo:art53,curtarolo:art55,curtarolo:art56,curtarolo:art57,curtarolo:art54,curtarolo:art63,curtarolo:art67,curtarolo:art70}. 
When identified, they are also added to the structure database.

The \aflow\
database also includes a few million bcc- and fcc-derived superstructures (those containing up to 20 atoms/cell), and a similar number of 
hcp-derived superstructures (all those containing up to 24 atoms/cell) enumerated formally using the algorithm of
Refs.~\cite{enum1} and \cite{enum2} (a derivative superstructure is a structure whose lattice vectors are
multiples of those of a parent lattice and whose atomic basis vectors correspond to \emph{lattice points} of the parent lattice).
In practice only a few hundred of these structures are usually calculated for each alloy system, but they are useful in
constructing cluster expansions
 \cite{Sanchez_ducastelle_gratias:psca_1984_ce,deFontaine_ssp_1994,gus:nature_evolutionary,uncle} which can be employed
in synergy with the HT calculations to examine structural
configurations more thoroughly than is possible with a purely HT
approach \cite{curtarolo:art49}.

\section{Basic operation}

At the beginning of a calculation, the starting structure, or
structures, are selected
from the database. The database structural description contains the
lattice parameters and atomic positions of the prototype
compound. \aflow\ then adjusts the lattice parameters (using a weighted average of the pure element volumes) for the specified elements and creates an input file
with all the necessary parameters for relaxations, static and band structure runs.
This entire procedure requires only the
label of a structure in the \aflow\ database. It may be performed with equal ease for only one structure with a specific composition or for a large set of structures
with many combinations of elements, with a single command. For multiple structures or
compositions this command creates a set of subfolders, each
including a single input file, through which
\aflow\ runs automatically until all structures and compositions are calculated.

A HT computational framework must contain a general, reliable, and standardized electronic structure
analysis feature. For example, it must automatically determine the Brillouin Zone (BZ) integration path for the
14 Bravais Lattices (with their 24 Brillouin zones) \cite{aflowBZ} and change the basis (lattice vectors) into a
standardized basis so that data can be compared consistently between different projects. Although BZ
integration paths have appeared in the literature in the last few decades
 \cite{Bradley72,Burns90,Miller67,Kovalev65,Casher69}, a standardized definition of the paths for all
the different cases has been, to the best of our knowledge, missing, and mistakes in the literature for
less-common Brillouin zones are not uncommon. This component of \aflow\ was discussed in detail in a separate
publication \cite{aflowBZ}.

\aflow\ computes structure total energies and electronic band structures using
\vasp\ with pseudotentials and accuracy chosen by the user.
As a default, \aflow\ employs projector augmented wave (PAW) pseudopotentials with GGA exchange correlation functionals \cite{DFT,PAW} as parameterized
by Perdew-Burke-Ernzerhof \cite{PBE}. Each structure is fully relaxed twice
with a convergence tolerance of 1 meV/atom using dense grids
of 3,000--6,000 $\kk$-points per reciprocal atom.
At the beginning of these structural relaxation steps, a spin-polarized calculation is performed for all structures.
If the magnetization is smaller than 0.025 $\mu_B$/atom, the spin is
turned off for the next relaxations and subsequent calculations to
enhance the calculation speed. After this, the structure is changed into a
standard form \cite{aflowBZ}, and another electronic relaxation is
performed using fixed coordinates for the lattice vectors and atomic
positions. This static run is implemented with a much denser grid of 10,000--30,000 points to get accurate charge
densities and density of states for the calculation of the band structure.
The Monkhorst-Pack scheme \cite{MonkhorstPack} is employed in the grid
generation except for hexagonal, fcc and rhombohedral systems in which
$\Gamma$-centered grid is selected for faster convergence.
This process is performed in one calculation, with the single input file \aflow\ creates as described above.
At the completion of such a calculation \aflow\ invokes appropriate \matlab\ or \gnuplot\ scripts for data analysis and
visualization. All these steps are usually performed automatically,
but the user also has the option to operate them through the web interface.
An example of a computed electronic band structure and standard BZ path appears in Figure \ref{fig1}.

\aflow's capability to continuously search subfolders for
calculations to run is not limited to DFT calculations. An {\it ``alien''} mode is also implemented,
which allows \aflow\ to execute other tasks in a high throughput fashion. For instance, the many
thousands of grand canonical Monte Carlo calculations used in a recent
surface science absorption project
 \cite{curtarolo:art44,curtarolo:art43} were directed and performed by
\aflow. In addition, \aflow\ is equipped with options to run
commands or scripts before and after the main program is
performed in each folder. This allows \aflow\ to generate input files
on the fly depending on the results of different calculations, so that
ad-hoc optimization can be implemented by the user.
It also improves the flexibility of recovery from a crash or an unconverged run and
increases the overall versatility and throughput of the calculation.

\section{Structure analysis tools}
\label{SAT}

\begin{figure}
\includegraphics[width=\linewidth]{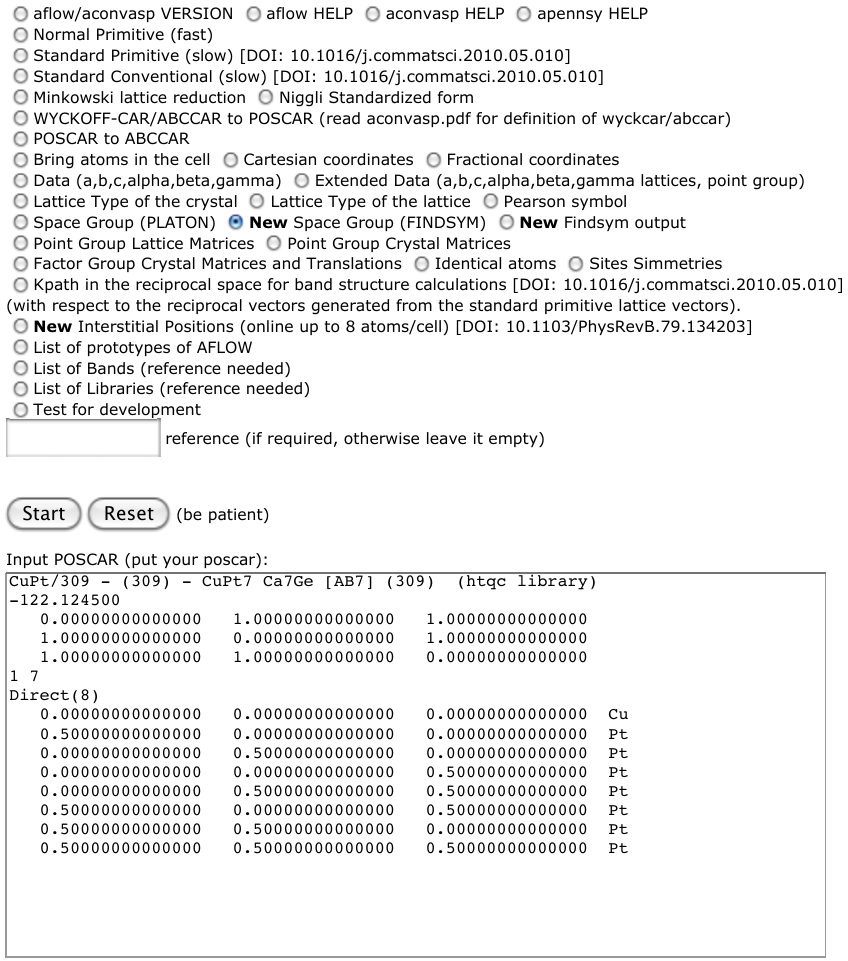}
\caption{Input interface for the structure analysis tools of \aflow. Users may apply these
 tools to a structure included in the database or enter their own structure in the input box. The
 tools include symmetry analysis, format conversions, and transforming structures between
 equivalent representations (primitive versus conventional, etc.). \label{fig2}}
\end{figure}

\aflow\ offers structure analysis and manipulation tools
which are also useful to users who do not need to perform HT calculations
or to create databases for datamining.
These users may prepare standard unit cell input
files and extract the appropriate {\bf k}-points path using the
command version of \aflow, called \aconvasp, or the
online interface on our website {\sf aflowlib.org/awrapper.html},
(shown in Figure \ref{fig2}).
Runs should then be performed according to the following protocol:
Unit cells must first be reduced to standard primitive, then
appropriately relaxed by \aflow. Before the static
run, the cells should be reduced again to standard primitive, since symmetry
and orientation might have changed during the relaxation. The user
should then perform the static run and then project the eigenvalues
along the directions which are specified in the ``k-path'' option. If
the user is running \aflow\ and \vasp, the web interface can also
prepare the input file, {\sf aflow.in}, which will perform all the
mentioned tasks.
The conversion and analysis operations may be carried on a structure file stored in
the database, or supplied by the user in the input box {\sffamily Input POSCAR}.
The operations implemented on the web interface are the following:
\begin{itemize}
\item {\sffamily Normal Primitive};
  Generates a primitive cell in the most compact form (not necessarily unique).
\item {\sffamily Standard Primitive}; 
  Chooses a primitive cell such that the Wigner-Seitz cell defined by the reciprocal vectors coincides with one of the possible 24 Brillouin zones \cite{Burns90,Bradley72,aflowBZ}.
\item {\sffamily Standard Conventional};
  Generates a conventional unit cell (not necessarily primitive).
\item {\sffamily Minkowski lattice reduction}; 
  Generates a maximally-compact cell (not necessarily unique).
\item {\sffamily Niggli Standardized form}; 
  Generates a cell that conforms to the Niggli standard form \cite{Niggli1928}.
\item {\sffamily WYCKOFF-CAR/ABCCAR to POSCAR}; 
  Generates a POSCAR file from a structure file containing the standard crystallographic information.
\item {\sffamily POSCAR to ABCCAR}; 
  Generates a structure file containing lattice parameters (rather than lattice vectors) from a POSCAR file.
\item {\sffamily Bring atoms in the cell}; 
  Remaps atomic coordinates to lie inside the unit cell.
\item {\sffamily Cartesian coordinates}; 
  Converts the atomic coordinates in a structure file to Cartesian coordinates (from lattice coordinates).
\item {\sffamily Fractional coordinates}; 
  Converts atomic positions from fractional coordinates (i.e, lattice or direct coordinates) to Cartesian coordinates.
\item {\sffamily Data} and {\sffamily Extended Data}; 
  Generates lattice parameters (cell lengths and angles) and other information (volume, reciprocal lattice, symmetry information, etc.).
\item {\sffamily Symmetry Information};
  Lattice Type of the crystal and the lattice, Pearson symbol, Space group and Wyckoff positions, Point Group Lattice Matrices, Point Group Crystal Matrices and factor Group Crystal Matrices and Translations.
\item {\sffamily Identical atoms and site symmetries};
  Identifies which atoms in a structure correspond to the same Wyckoff positions.
\item {\sffamily K-path in the reciprocal space}; 
  Provides directions in {\bf k}-space needed for band structure calculations and provides a picture of the corresponding Brillouin zone and path.
\item {\sffamily Interstitial Cages Positions}; 
  provides all the geometric locations of interstitials in the structure.
\end{itemize}

The user interface also provides access to the software manuals, under
the buttons marked {\sffamily aflow HELP, aconvasp HELP} and
{\sffamily apennsy HELP}, and to the current list of experimental
prototypes in the \aflow\ database under {\sffamily List of prototypes of AFLOW}.

\begin{figure*}
\parbox{.75\linewidth}{
\includegraphics[width=.9\linewidth]{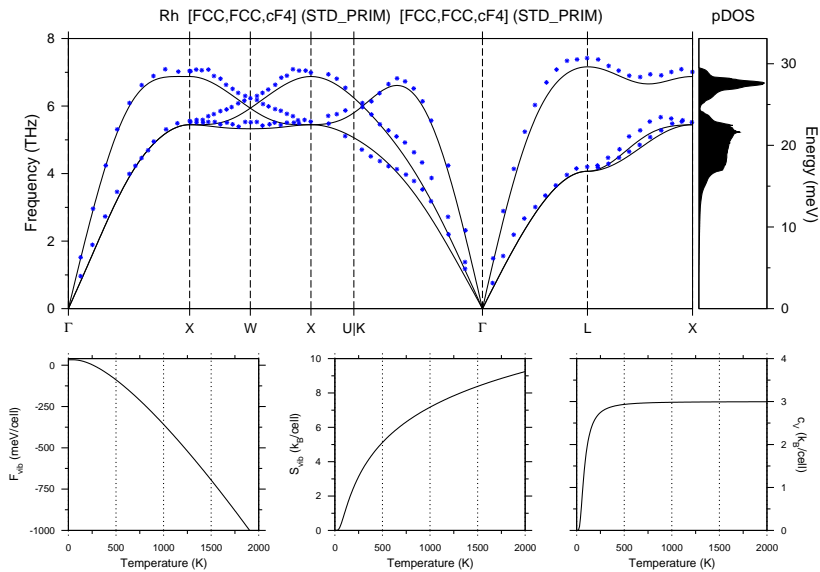}
}
\parbox{.22\linewidth}{ \caption{ Top: Phonon dispersion curves for face-centered cubic
  rhodium. Solid lines are computed frequencies, dots are experimental results
  \cite{EichlerPRB98}. The phonon density of states is shown on the right hand side. Bottom: Vibrational free energy, vibrational
  entropy, and specific heat calculated automatically using the \aflow\
  framework. \label{fig3}}
 }
\end{figure*}

\begin{figure}[ht]
\includegraphics[width=0.4\textwidth]{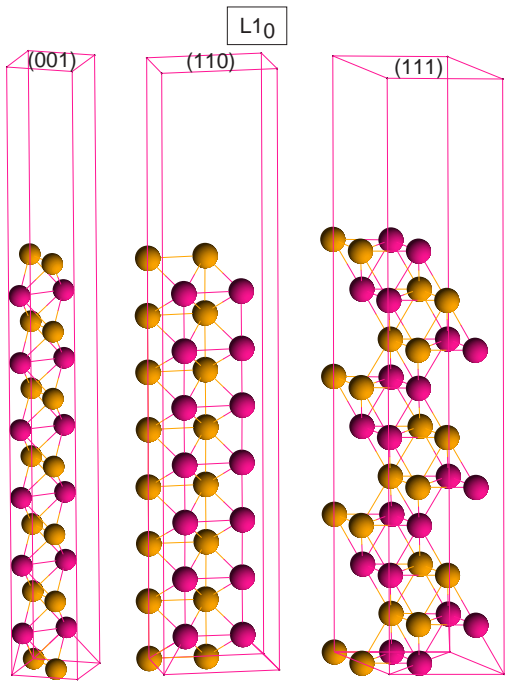}
\caption{The supercells of (001), (110), and (111) slabs of fcc
binary L$1_0$ structure built by \aconvasp.} \label{fig4}
\end{figure}

\section{Example Applications}

\subsection{Vibration spectra and free energy}

\aflow\ can calculate phonon dispersion curves using three different approaches:
the direct force constant method, following the definition of Ref.~ \cite{Maradudin,Madelung} and description of Ref.~ \cite{Kresse1995729},
the linear response method for PAWs \cite{Gajdos2006}, and
the frozen phonon method as implemented in the \frozsl\ code of H. Stokes and L. Boyer \cite{Stokes_FROZSL_Ferroelectrics_1995,Boyer_Stokes_Mehl_Ferroelectrics_1995,FROZSL_website}.

In the first method, the atomic displacements necessary to fully determine the dynamical matrix are prescribed automatically, the forces
are calculated by \vasp\ and mapped onto symmetry-equivalent directions by \aflow. 
The method is general and can treat low-symmetry structures as easily as high-symmetry systems, minimizing user-time overhead.
To reduce computational efforts, the symmetry of the supercells is
fully employed and only the unique atoms are distorted along the
minimum independent directions required for the construction of the force
constant matrix. When necessary, we use linear response \cite{Gajdos2006} for the non-analytical part of the dynamical matrix
\cite{Gonze1997,Wang2010}, to reproduce the correct LO-TO splitting (splitting between longitudinal and
transverse optical phonon frequencies). 
For the linear response method as implemented in \vasp, \aflow\ sets
up the input files, runs and converges the calculation automatically.
\aflow\ uses the generated dispersion curves to calculate the vibrational
contribution to the free energy, entropy and specific heat (see example in Figure \ref{fig3}).
The whole task is performed by the {\small APL} library of \aflow.

In the frozen-phonon approach, \aflow-\frozsl, the user sets up an {\sf aflow.in} input file containing instructions in the \frozsl\
format \cite{FROZSL_website}.
To perform the calculation \aflow\ repeatedly calls \frozsl\ to generate
all the irreducible representations of the phonons, calculate the
various geometries and the phonon spectrum in the requested {\bf k} points or paths.
The code is multi-threaded: the various irreducible representations can be tackled simultaneously in a computer cluster 
while an \aflow\ daemon waits for their completion to produce the spectrum. 
For instance, the several thousand calculations of Ref. \cite{curtarolo:art60} were rapidly addressed by this method.

We chose to implement all three approaches for convenience. 
While the first gives more information in the whole Brillouin Zone, the second and the third can be used to rapidly scan
spectra at particular {\bf k} points, such as when searching for Kohn anomalies \cite{curtarolo:art37}.
Currently, the \frozsl/\findsym\ source packages are maintained by the \aflow\ team. 
They are included in the \aflow\ distribution and have been modified to compile with the GNU suite.

\subsection{Design of high-index surfaces in complex multicomponent compounds}

Surface segregation, adsorption, chemisorption, catalytic reaction and other surface phenomena are of ultimate importance in many areas
of modern technology, including catalysis, corrosion protection, batteries, electronics etc. In most theoretical studies of surface effects, it
is necessary to know the exact positions of surface and subsurface atoms. In cases of low-index surfaces and simple compounds, the
construction of a surface can be done manually. However, the technologically interesting high-index
surfaces \cite{Tian07,curtarolo:art61} and multicomponent compounds with complex underlying crystal lattices require an
automated tool.

The input file used for the surface construction describes the initial bulk crystal structure, denoted by the corresponding \aflow\ database label. The structure may have an
arbitrary complex crystal lattice with any number of atom types. The generated surface file contains the complete data (both direct and Cartesian
coordinates) for the constructed surface supercell. The Miller indices $(hkl)$ are based on the Bravais lattice that corresponds to the unit cell of the bulk crystal structure.
An example of (001), (110), and (111) supercell slabs of the fcc binary L$1_0$ structure are shown in Figure \ref{fig4}.
\aflow\ can easily build a supercell containing a predetermined number $N_f$ of atomic planes with given Miller indices $(hkl)$ of a given crystal structure. A supercell may also
contain a number $N_e$ of ``empty'' $(hkl)$-planes, without actual atoms. In the last case, a periodically repeated supercell
represents the slabs separated by empty space. Such slab construction allows one to study surface effects, e.g. surface
energy, surface stress, and surface segregation, from first principles \cite{Fiorentini96,Ruban99,Ruban09,Marzari09,curtarolo:art50,curtarolo:art61}.
The obtained surface data may also be effectively used in simulation of nanoparticles \cite{curtarolo:art50,curtarolo:art61}.
The variation of $N_f$ and $N_e$ is important for studying convergence with respect to slab width and inter slab distance, respectively.
Within $(hkl)$-planes, the constructed supercell consists of $p_1 \times p_2$ two-dimensional primitive unit cells, where $p_1$, $p_2 =
1,2,\ldots$ are specified. Large $p_1$ and $p_2$ are used to introduce surface defects, e.g. vacancy or solute atoms, far enough from each other to diminish
their interaction. For instance, all the surfaces studied in Ref. \cite{curtarolo:art61} were produced with this method.

\subsection{Nanoparticle generation}

\begin{figure}[ht]
\includegraphics[width=0.95\linewidth]{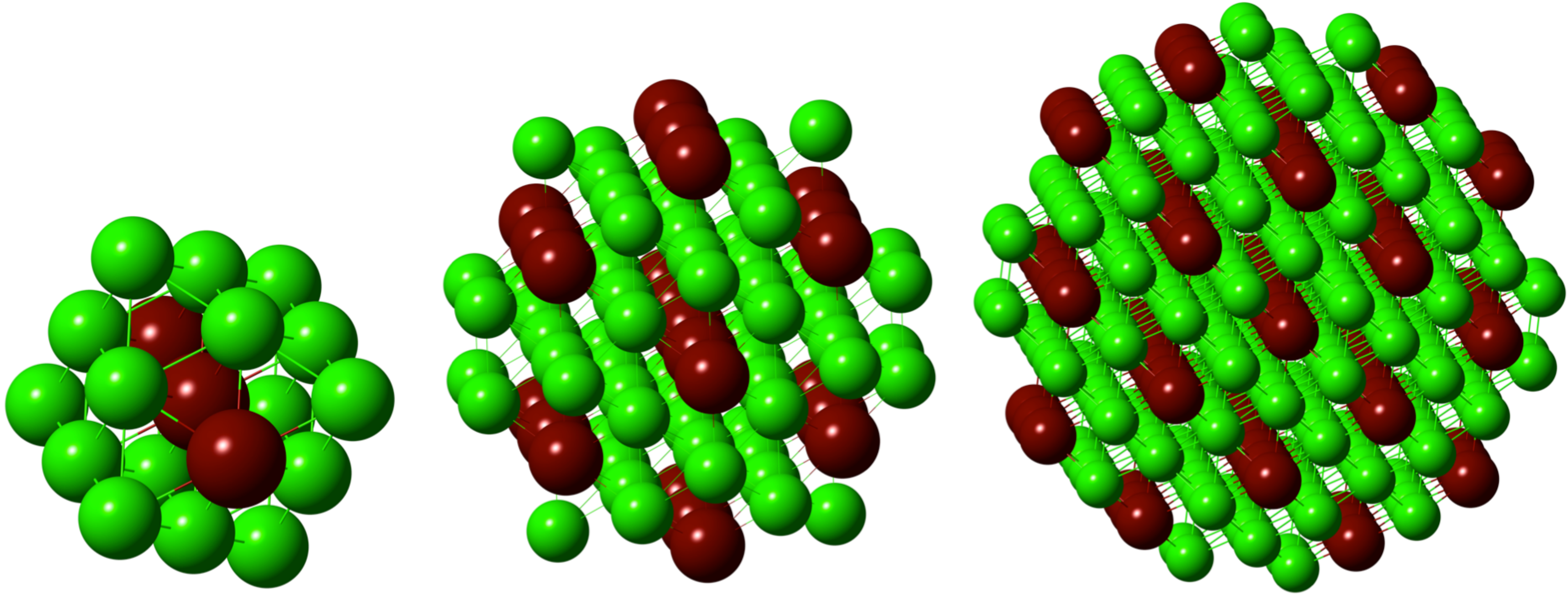}
\caption{Nano-particles based on the L$1_3$ structure built by \aconvasp, with radii of 5\AA\ (left), 8\AA\ (center) and 12\AA\ (right).}
\label{fig5}
\end{figure}

Starting from an input crystal structure, a specified radius and required separation, \aconvasp\ generates a structure file of a nano-particle of that radius made of the same lattice and
separated as specified from its nearest neighbors. The origin of the particle may be set to the origin of the unit cell, an atom in the crystal structure or any point in Cartesian or fractional coordinates.
The radius and separation distance are in \AA. This option replaces the cumbersome manual generation of nano-particle structure files, usually employed in studies of such
particles, and should enable investigation of large sets of nano-particles in a high-throughput fashion. Figure \ref{fig5} shows examples of the nanoparticle generation.

\subsection{Topological identification of interstitial sites}

An additional feature of \aflow\ allows users to identify interstitial sites inside any crystallographic structure. Using as input a file specifying the atomic positions in the structure
unit cell, it returns the location, radius and coordination of each site. The algorithm identifies quadruplets of non-coplanar atoms, where the first atom belongs to the unit cell and the
others are separated by less than the longest diagonal of the unit cell. {\it Cages} are defined by spheres touching all four atoms of the quadruplet that do not contain any atoms.
An {\it interstitial} position is found if a center of a cage is inside the unit cell. By considering all the possible combinations, symmetrically inequivalent
interstitials can be identified through calculation of their site symmetry (with the space group of the unit cell). Note that in unit cells with complex arrangements, many of the interstitial positions can be extremely close.
Thus, adjacent interstitial atoms located in any of those close positions could deform the nearby local atomic environment and
{\it relax} to the same final location. Hence, the number of symmetrically inequivalent cages can be further reduced by {\it agglomeration} of
such a set of positions into a single interstitial site, upon insertion of interstitial atoms.

\begin{figure}[htb]
\includegraphics[width=0.70\linewidth]{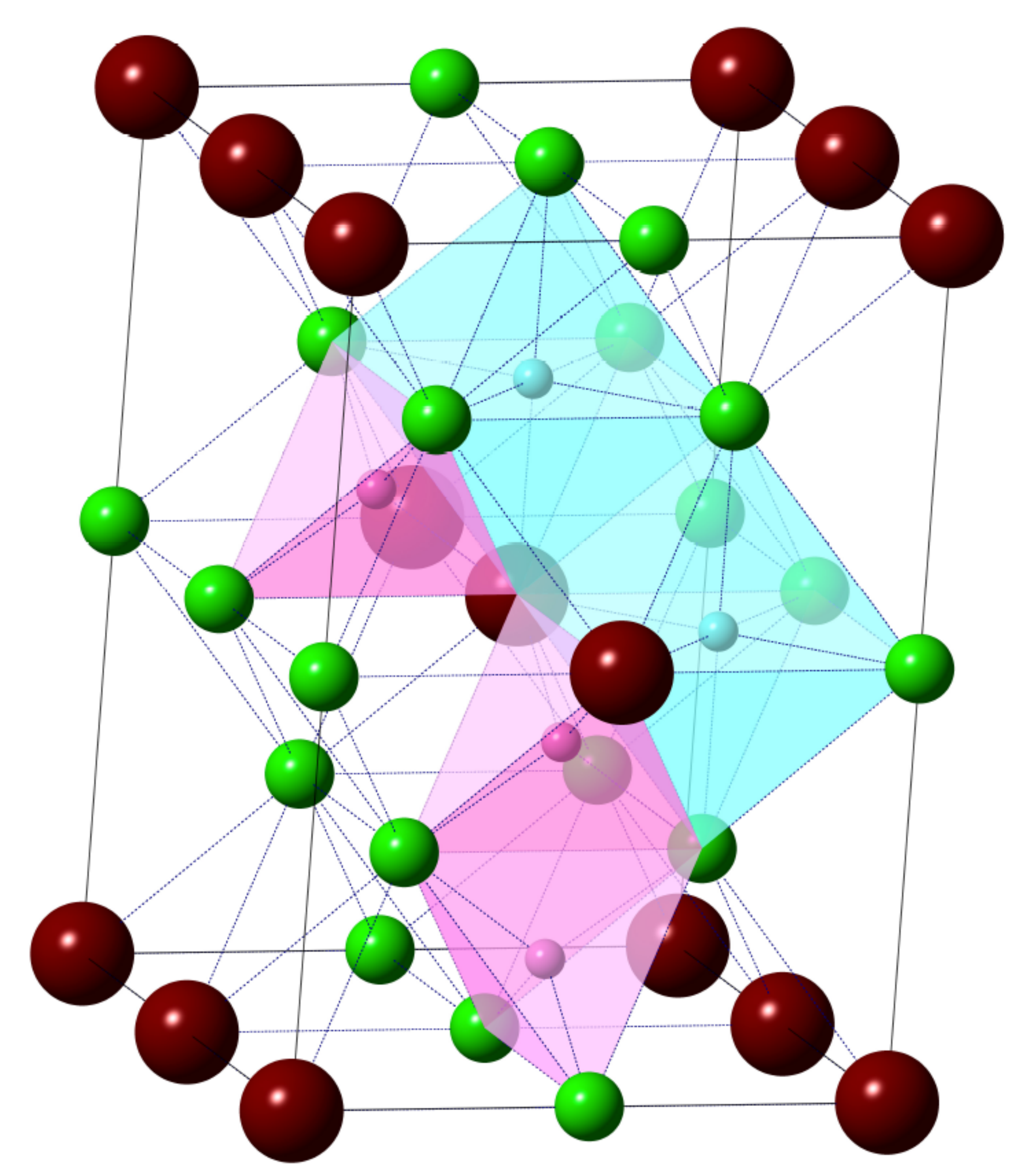}
\caption{6-coordinated (blue) and 4-coordinated (pink) interstitial
sites in the L$1_3$ structure, found in a topological search by \aconvasp.}
\label{fig6}
\end{figure}

\begin{figure}[h!]
\includegraphics[width=0.95\linewidth]{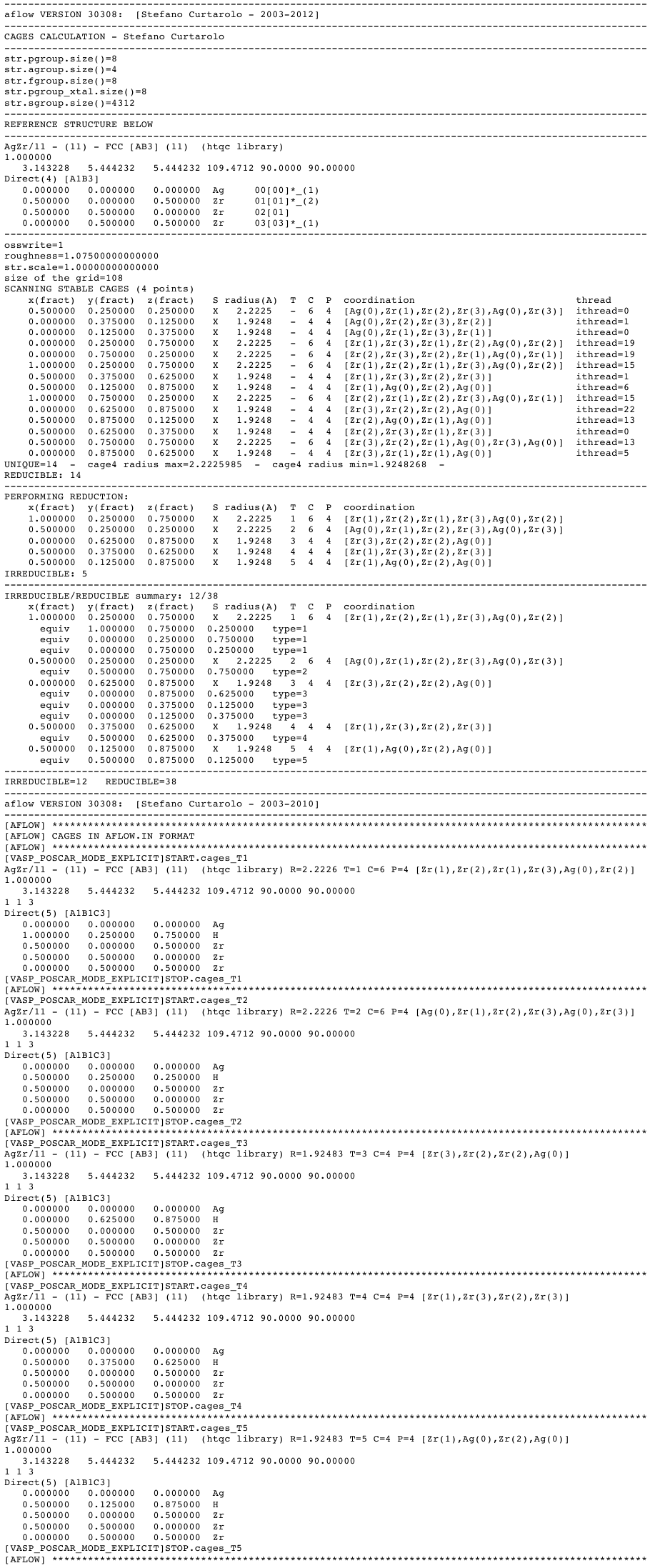}
\caption{Output of a multi-threaded search of interstitial cages in the L$1_3$ structure, showing two octahedral positions with $r=2.2225$\AA
(Ag$_1$Zr$_5$ and Ag$_2$Zr$_4$) and three tetrahedral positions with $r=1.9248$\AA\ (Ag$_1$Zr$_3$, Ag$_2$Zr$_2$ and Zr$_4$). 
The cell geometries including these interstitials are given in the bottom part of the printout.} 
\label{fig7}
\end{figure}

Given the host cell geometry and the interstitial species, the occupation of the final irreducible cages can be automatically simulated by \aflow, which 
calculates their energies, entropy and solubility in the small concentration limit \cite{curtarolo:art41,curtarolo:art47}.
These operations are implemented in a multi-threaded manner to accelerate the calculation in a multi-core environment.
Figure \ref{fig7} presents an example of the output of a multi-threaded interstitial position search performed online through our servers ({\sf aflowlib.org/awrapper.html}), for
the AgZr$_3$-L1$_3$ structure, using the interface of Figure \ref{fig2}.  Here \aflow\ finds 14 cages: 6 octahedral and 8 tetrahedral.
Reduction through the space group leads to only 5 irreducible positions, two octahedral positions with $r=2.2225$
\AA\ (Ag$_1$Zr$_5$ and Ag$_2$Zr$_4$) and three tetrahedral positions with $r=1.9248$ \AA\ (Ag$_1$Zr$_3$, Ag$_2$Zr$_2$
and Zr$_4$). The equivalence and the multiplicity per cell are provided in the text.
Figure \ref{fig6} shows these octahedral and tetrahedral interstitials in the standard conventional representation of this
structure L1$_3$. For clarity, the conventional oP8 structure is shown, instead of the primitive oS8. The two unique octahedrons are blue and the three unique tetrahedra are in
pink. The interstitial atom is in the center of the polytopes.

\subsection{\apennsy: automatic analysis}

The results of high-throughput \aflow\ runs are analyzed and manipulated for further processing
by the \apennsy\ module. The basic operation of this module is the production of a system-specific
data file that includes information
on all the computed structures, their energies, the input and output (after relaxation) crystal prototype and space group,
and the binary convex hull. It requires as input the output files generated by the \aflow\ run for the computed system.

For example, in a binary alloy system, the {\it convex hull} at zero-temperature is used to create the phase diagram.
To construct the convex hull, \apennsy\ finds the minimum energy structure at each composition,
and calculates their formation enthalpies
\begin{equation}
H_f=E(x)-xE_A-\left(1-x\right)E_B,
\label{H_f}
\end{equation}
where $E_A$ and $E_B$ are the energies of the pure elements and $E(x)$ is the energy of the alloy with concentration $x$
of element $A$. \apennsy\ then identifies the extremal structures, i.e., those that lie below the tie-lines
connecting their neighbors at adjacent compositions. These extremal structures are the stable structures in the
binary system and constitute its convex hull. \apennsy\ plots the convex hull automatically using \gnuplot\ or \matlab.

The module can also be used to print the computed data in formats appropriate for use as input for other codes.
For example, it can print all the bcc (or fcc or hcp)-based structures with the total energy per unit cell or energy per atom.
This is useful for cluster expansion codes such as \uncle\ \cite{uncle}. Another format is used for automatic miscibility determination.
All these data files can be generated off-line by operating \apennsy\ on a library of input files, or
on-line by selecting a specific system and the requested option at the \aflow\
website {\sf aflowlib.org} (``binary alloy library'' entry) \cite{aflowlibPAPER}.
Additional options of \apennsy\ provide data that is external to the structure calculations, such as predictions of system
miscibility following Miedema and Hume-Rothery mixing rules \cite{Miedema,hume_rothery}.

\section{Future developments: High-throughput hybrid functionals calculations}
\label{Demchenko stuff}

The local (LDA) and semi-local (GGA) approximations to the DFT, expanded with ``+U'' techniques \cite{LDAU},
have been incredibly successful in describing a wide variety of 
materials properties, particularly those related to the ground state. In some cases, however, the drawbacks of these approximations 
present an insurmountable obstacle. 
One of the most illustrative areas is the energetics of defects in wide band-gap semiconductors.
For example, in the case of oxygen vacancies in ZnO, the defect formation energies and thermodynamic transition levels 
obtained by different LDA/GGA approaches vary by several eV \cite{Lany}.
Similar problems exist in DFT calculations of optical properties of insulators and semiconductors, properties of correlated materials, 
reaction energetics and adsorption of small molecules \cite{reaction,ads_molecule}.
An affordable and accurate approach to overcome these problems does not exist yet. 
The state of the art many-body methods, such as self-consistent GW \cite{Kresse_GW}
exhibit excellent agreement with experiments, but suffer from high computational cost. 
The hybrid functional methods\cite{Kresse_hybrid1} represent a pragmatic compromise between the local and nonlocal many-body approaches. 
The results of hybrid functional calculations are often in excellent agreement with computationally expensive cutting edge many-body methods, 
yet favorable scaling allows routine calculations of a few hundred atoms in the unit cell \cite{Kresse_hybrid2} .

In a hybrid functional calculation the LDA exchange correlation part of the density functional is mixed with a 
Fock-type exchange part in varying proportions. For example, the popular Perdew-Burke-Ernzerhof hybrid functional (PBE0) 
contains 25\% of the exact exchange, 75\% of GGA exchange, and 100\% of GGA correlation energy \cite{PBE0}. 
In a related Heyd-Scuseria-Ernzerhof (HSE06) functional the Fock exchange interactions are separated into long- and short-range parts. 
The short range part includes 25\% of exact
exchange and 75\% of semi-local GGA exchange, while the long-ranged part is replaced with an approximate semi-local GGA expression \cite{HSE06}.
The splitting is accomplished by introducing the screening of the exchange interactions (similar to the screened exchange approach) 
with optimal screening length of approximately 7-10 \AA. 
This approach presents a middle case between the GGA calculation (no exact exchange) and the PBE0 hybrid functional (all exchange is long ranged). 
The HSE06 functional eliminates some unphysical features of the exact
exchange approach. It also has the computational advantage of better
convergence of the long-ranged exchange part, since screening out the long range exact exchange greatly reduces
the cost of calculating the non-local exchange interaction. 
One also often finds it useful to tune the amount of exact exchange
for a particular material (the so called $\alpha$-tuned hybrid functionals) in order to obtain the best agreement 
of computed band-gap with experiment.
This is particularly useful in cases where the correct value of the band-gap is critical. For example, in the calculations of defect energetics, 
tuning the band-gap of the host material in the bulk and then running calculations for supercells containing impurities and defects
has been shown to produce results in very good agreement with experiment \cite{Oba} .

The hybrid functional calculations are computationally more demanding than LDA/GGA calculations, due to the need to fully 
calculate the non-local Fock exchange integrals. For example, a bulk ZnO HSE06 calculation ($8\times8\times6$ $\Gamma$-centered
{\bf k}-point mesh, with 40 irreducible {\bf k}-points, 400 eV cutoff) using \vasp\ takes approximately 5 hours on 8 CPU's.
The same set up for the GGA/PBE (+U) calculation takes less than a minute. 
The memory requirements in this case, for a HSE06 calculation, are also increased by a factor of five compared to those of GGA. 
The scaling of the computational cost varies for different hybrid functionals, and is approximately $O(N^{2.5})$ for small system sizes, 
and linear for HSE06 beyond 15 \AA~ and for PBE0 beyond 100 \AA.
Using plane waves the scaling of hybrid functional methods is
$(N_{\mathrm{bands}}N_k)^2N_{\mathrm{FFT}}log(N_{\mathrm{FFT}})$, which is approximately linear with the number of atoms in the 
bulk \cite{Kresse_hybrid2}. The standard (semi-) local DFT(+U) methods using plane waves normally scale as $O(N^{3})$.

\subsection{Framework of DFT+U and Hybrid Functional calculations}

{\bf HT-DFT+U:} Within the high-throughput framework, \aflow\
currently employs the DFT+U technique and allows the user to choose the appropriate parameters.
As default, \aflow\ takes advantage of the Dudarev \cite{DudarevDFTU} formalism within GGA+U. The values of $U$ are listed in Table \ref{tab:Ueff}.

\begin{table}[!h]
 \caption{\small
  Values of $U_\mathrm{eff}$ parameters in \unit{eV} for the Dudarev GGA+U approach implemented in \aflow. From Refs. \cite{aflowSCINT,aflowBZ,aflowTHERMO} and references therein.}
 {\footnotesize
  \begin{tabular}{cccccccccc} \hline \hline 
   Ti & V & Cr & Mn & Fe & Co & Ni & Cu & Zn & Ga \\
   4.4&2.7&3.5&4.0&4.6&5.0&5.1&4.0&7.5&3.9 \\ \hline 
   & Nb & Mo & Tc & Ru & Rh & Pd & & Cd & In \\
   &2.1&2.4&2.7&3.0&3.3&3.6&&2.1&1.9 \\ \hline 
   & Ta & W & Re & Os & Ir & Pt & & &\\
   &2.0&2.2&2.4&2.6&2.8&3.0&&& \\ \hline \hline
   La & Ce & Pr & Gd & Nd & Sm & Eu & Tm & Yb & Lu \\
   7.5&6.3&5.5&6.0&6.2&6.4&5.4&6.0&6.3&3.8 \\ \hline
  \end{tabular}
 }
 \label{tab:Ueff}
\end{table}

{\bf HT-DFT+Hybrid:} We are currently expanding the \aflow\ framework to perform HSE06 calculations, and we present some preliminary results.
The extension is not trivial: the calculations of electronic structure using hybrid functionals differ from the standard LDA/GGA ones, with or without ``+U''. 
One cannot compute the bands in the familiar non-selfconsistent way of the LDA/GGA because non-local exchange is not determined by the pre-computed charge density. 
Therefore to obtain the eigenvalues for strings of {\bf k}-points the following recipe is being implemented.
First, \aflow\ performs a standard LDA/GGA calculation.
Second, a hybrid functional run is performed starting from the LDA/GGA wavefunctions on the same {\bf k}-point mesh and energy cutoff chosen by \aflow.
The number of bands is dynamically adjusted to achieve full convergence. 
Then, to facilitate efficient use of computational resources, it is
further adjusted to include only a few bands over the highest occupied state. 
Third, the electronic structure is computed by performing a hybrid functional run explicitly defining the same {\bf k}-point mesh in addition to the desired {\bf k}-path
as specified in the high symmetry examples of Ref. \cite{aflowBZ}. 
The crucial step here is to set the mixing weights of the extra {\bf k}-path to zero, while keeping the original mesh intact. 
Since the orbitals at the extra {\bf k}-path do not contribute to the total energy, and the wavefunctions on the original mesh are converged as input,
it is only necessary to converge the orbitals at the extra {\bf k}-points mesh with the appropriate \vasp\ instructions.
The band structure calculation following this recipe will take approximately the same amount of time as a regular total energy
hybrid calculation since the orbitals at the standard mesh are pre-converged. 

\begin{figure}[!h]
 \includegraphics[width=0.99\linewidth]{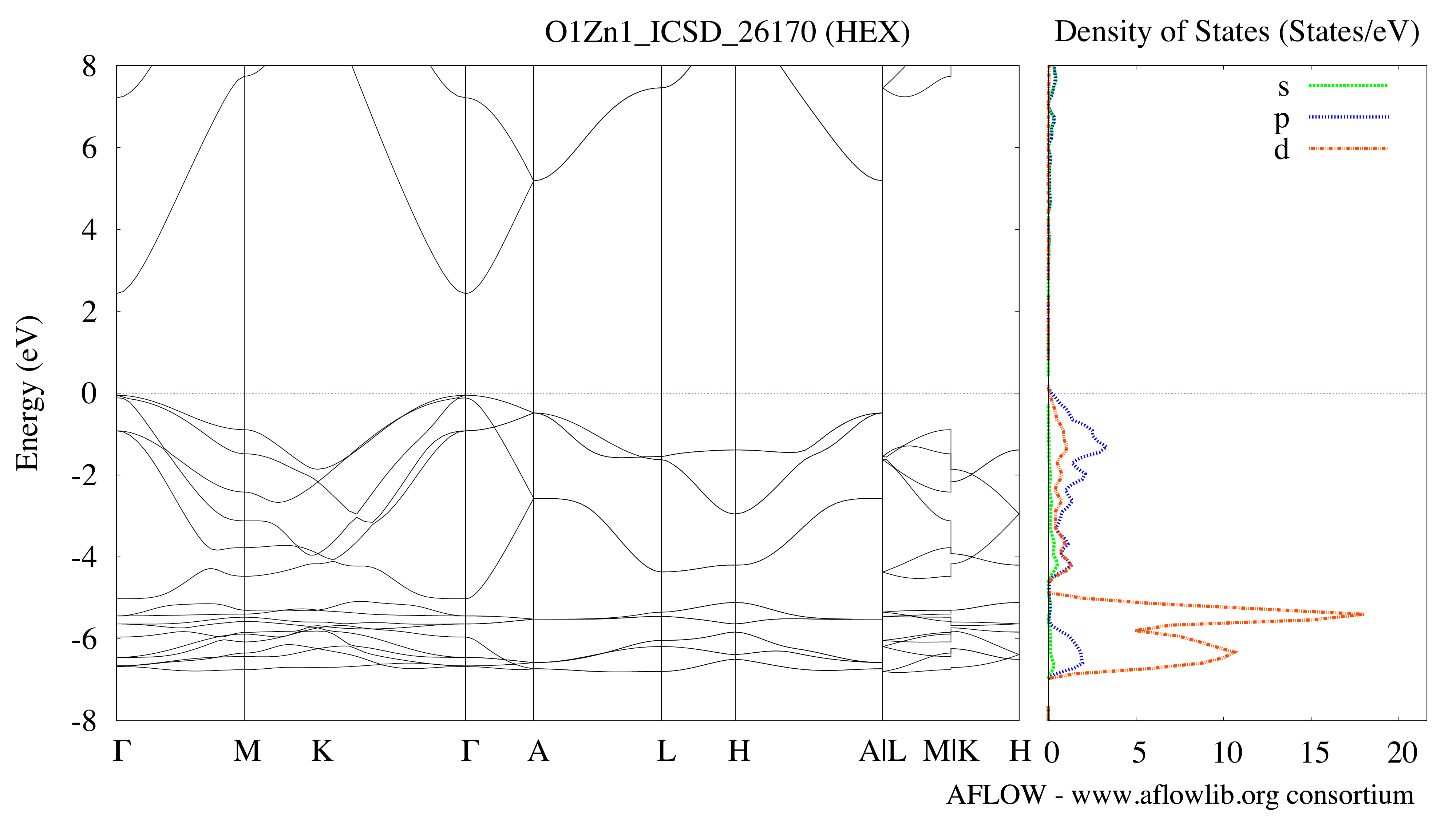}
 \caption{HSE06 hybrid functional band structure of ZnO.
  The HSE06 gap is 2.48 eV. 
  LDA/GGA gap is 0.7 eV while the PBE+U value is 1.82 eV \cite{aflowlib.org.O1Zn1_ICSD_26170}.
  Standard HSE06 parameters were used, the ratio of exact exchange is 0.25 and the screening parameter is 0.2 \AA$^{-1}$. 
  \label{fig8}}
\end{figure}

As an example Figure \ref{fig8} shows the HSE06 band structure computed along the standard high symmetry lines. 
Comparisons can be drawn with the PBE+U electronic structure of Ref. \cite{aflowlib.org.O1Zn1_ICSD_26170}.
The HSE06 calculation used the standard value of $\alpha=0.25$. (Note that by using $\alpha=0.375$ the band-gap can be
brought to agreement with experiment.) The value of the band-gap is considerably improved $E_g=2.48$ eV in comparison with
GGA and PBE+U values of $E_g=0.7$ eV and $E_g=1.82$ eV, respectively (the phenomenologically corrected value is 3.36 eV \cite{aflowSCINT}).
In other materials this improvement is often better, since ZnO produces one of the largest LDA/GGA band-gap errors. 
The HSE06 band-gaps for narrow and medium gap semiconductors, with the standard ratio of the Fock exchange, are in good agreement with experiment. 
For wide band-gap semiconductors and insulators the gaps, although underestimated, present 
a significant improvement over LDA/GGA values \cite{Kresse_hybrid1}.
In addition, hybrid functionals improve effective mass estimation for almost all systems,
typically yielding values within a few percent of experiment \cite{hybrid_eff_mass}.

A fully functional, consistent and robust \aflow\ framework with hybrid functionals is planned for 2012.

\section{Summary}
We describe the \aflow\ software package for HT calculations of material properties.
It should be helpful to the materials scientist seeking to determine properties of alloy and compound structures.
The code, and operation manuals describing its features, are freely
available for download at {\sf aflowlib.org/aflow.html}.
The structure manipulation and analysis capabilities of \aflow\ are
also available online at the \aconvasp-online entry of this
 website.

\section{Acknowledgments} 
The authors acknowledge Gerbrand Ceder, Marco Buongiorno-Nardelli, Leeor Kronik, Natalio Mingo, and Stefano Sanvito for fruitful discussions.
Research supported by 
ONR (N00014-11-1-0136, N00014-10-1-0436, N00014-09-1-0921), NSF (DMR-0639822, DMR-0908753), and the
Department of Homeland Security - Domestic Nuclear Detection Office.
We are grateful for extensive use of the Fulton Supercomputer Center at Brigham Young University and Teragrid resources (MCA-07S005). 
SC acknowledges support by the Feinberg fellowship at the Weizmann Institute of Science.

\end{document}